\documentclass[twocolumn]{article}
\usepackage[papersize={8.27654in,11in},total={7.25in,10.1in}, left=0.55in, nohead,nofoot]{geometry}
\usepackage{txfonts}
\usepackage[colorlinks=true,allcolors=blue,bookmarks=false,
  pdfauthor={Gavin R. Putland},
  pdftitle={The cycloid as brachistochrone: A one-page proof, from first principles, without calculus},
  pdfkeywords={brachistochrone, Fermat's principle, refraction},
  pdfsubject={Geometry, Dynamics, Optimization}]{hyperref}
\usepackage{graphicx,tikz}
  \definecolor{myc}{rgb}{0,0.5,1} 
\usepackage[final,kerning,spacing]{microtype}
  \SetExtraKerning{encoding={*}}{1={-35,-45}, / ={40,50}, ( ={,-30}, ) ={-30,},
    \textemdash={50,50}, \textendash={30,30}, \textquotedblleft={-30,}}
  \SetExtraSpacing{encoding={*}}{A={-80,-80,-80}, r={-80,-80,-80},
    v={-80,-80,-80}, w ={-70,-70,-70}, y ={-60,-60,-60},
    \textquotedblright={-80,-80,-80}} 
  \SetExtraSpacing{font={*/*/*/n/*,*/*/*/sl/*,*/*/*/sc/*}}{y={-80,-80,-80}}
\title{\vspace{-3ex}{\Huge\bf The cycloid as brachistochrone:\LARGE\bf\\A~one-page proof, from first principles, without calculus}\vspace{-.8ex}}
\author{Gavin R.\,Putland~\!\thanks{\footnotesize\sf\,Royal Melbourne Institute of Technology, Australia.~ License: \href{https://creativecommons.org/licenses/by/4.0/legalcode}{Creative Commons Attribution 4.0 International}.}~~ \normalsize(version~2;\, November~4, 2025)}
\date{\vspace{-5ex}}

\begin{document}

\maketitle
\thispagestyle{empty}

\linespread{0.97}\selectfont

{\small\noindent\textbf{Abstract:}~ Johann Bernoulli's optical
solution of the brachistochrone problem is rebuilt on underlying
(non-optical) principles. An ``optical interpretation'' is given
afterwards.}

\begin{figure}[h]\centering
\vspace{-1ex}
\includegraphics{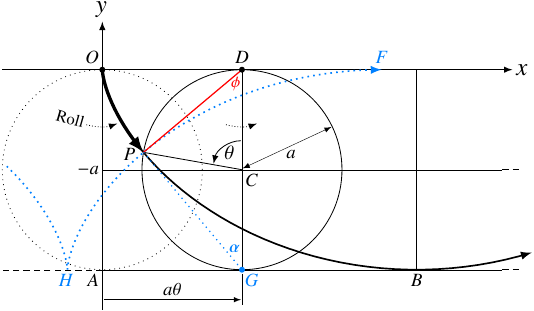}
\vspace{-3ex}
\end{figure}

\noindent A {\bf cycloid} is the plane curve traced by a point ({\bf
  generating point}) on a circle ({\bf generating circle}) which
rolls, without slip, along a fixed line ({\bf generating line}). If
the line is horizontal with the circle hanging vertically below~it,
the cycloid is ``{\bf inverted}''\!.\, In~the diagram, a~circle with
radius~$a\:\!$ has rolled along segment~\textit{OD} of the $x$-axis,
turning through angle~$\theta$ (radians) and traveling
distance~$a\theta$, so that the generating point~$P$, starting
from~$O$, has traced the arc~\textit{OP} of the inverted
cycloid~\textit{OPB}.

\vspace{-2ex}
\paragraph{\color{red}Normal chord:}
The infinitesimal piece of the circle at $D$~\! has no velocity
component in the $x$~direction (no~slip) or in the $y$~direction
(because it has reached maximum~\textit{y}): its instantaneous velocity is
zero. Hence the motion of $P$ must be \emph{normal to PD} (or else the
said piece of the circle would have a velocity component
along~\textit{PD}). Thus \emph{PD is the normal to cycloid~OPB at~P};
and in general, \emph{the normal to a cycloid at a point is given by
the chord of the generating circle from that point (as~generating
point) to the point of contact with the generating line}.

\vspace{-2ex}
\paragraph{\color{myc}Normal cycloids (\& cycloidal cylinders):}
If a second circle, congruent to and superimposed on the first, rolls
along the bottom line~\textit{AB}, the points of that circle trace
cycloids, one of which, namely~\textit{HPF}, passes through~$P$, where
cycloid~\textit{HPF} has normal chord~\textit{PG}\, and
cycloid~\textit{OPB} has normal chord~\textit{PD}. As the normal
chords make a right angle with each other (angle in a semicircle),
\emph{the cycloids are normal to each other at~P}. The same applies to
later positions of $P$ and~\textit{HPF}, including (by~symmetry)
positions for~which the part of~\textit{HPF} left of~$H\:\!$
intersects the part of~\textit{OPB} right of~$B$. Thus
\emph{cycloid~OPB is an \textbf{orthogonal trajectory} to the family
of cycloids generated by circle~GPD and line~AB}\textemdash and to the
family of {\bf cycloidal cylinders} generated by the latter cycloids
if they move in the $z$~direction (normal to the $xy$ plane).

\vspace{-2ex}
\paragraph{Horizontal \& normal velocities:}
Let angle~\textit{PDC} be~$\color{red}\phi$~\! as shown. Let
cycloid~\textit{HPF} travel in the $x$~direction at speed $u$, and let
$v$ be its resulting \emph{normal} speed at~$P$ (measured
along~\textit{PG}{\tiny~\!}). Then\vspace{-1ex}
\begin{equation}\label{e-v-u}
  v/u = \cos\phi \,.
\end{equation}
At~$P$,~ ${y=-\textit{DP}\cos\phi}$.\, But ${\textit{DP}=2a\cos\phi}$
(from the right-angled triangle~\textit{DPG}). Back-substituting
for~\textit{DP}~\! we have, at~$P$,\vspace{-.5ex}
\begin{equation}\label{e-y-ph}
  y = -2a \cos^{2\!}\phi \,.\vspace{-.5ex}
\end{equation}
Eliminating $\cos\phi$ between eqs.\,(\ref{e-v-u}) and (\ref{e-y-ph})
gives\vspace{-1ex}
\begin{equation}\label{e-kine}
  \mbox{\large$\frac{\,1\,}{2}v^2 + \frac{\;u^2}{4a}\:\!y = 0$} \,.
  \vspace{-1.5ex}
\end{equation}

\vspace{-2ex}
\paragraph{Motion under gravity:}
The first term in eq.\,(\ref{e-kine}) is the kinetic energy, per unit
mass, of a particle with speed~$v$. In a uniform gravitational
field~$g$ in the $-y$~direction, the particle's gravitational
potential energy per unit mass is~$gy$ (relative to {\it y}~\!=~\!0).
Hence, if the particle is released from rest at ${y\!=\!0}$ and moves
under gravity at speed~$v$ along a frictionless path, conservation of
energy gives\vspace{-1ex}
\begin{equation}\label{e-energy}\textstyle
  \frac{\,1\,}{2}v^2 + gy = 0 \,.\vspace{-2ex}
\end{equation}

\vspace{-2ex}
\paragraph{Path of least time:}
Given an origin $O$~\! and a destination point~$Z$ (not illustrated),
with $Z$ no higher than~$O$, let us choose the radius~$a$ and the
$x$~direction so that cycloid~\textit{OPB} passes through~$Z$. Then
eq.\,(\ref{e-energy}) determines $v$ as a~\! \emph{function
of\,\,$y$}, say~$v(y)$, on~all frictionless paths from $O$
to~$Z$\textemdash including cycloid~\textit{OPB}\textemdash if the
particle is released from rest at~$O$. Now let the horizontal speed
$u$ of cycloidal cylinder~\textit{HPF} be\vspace{-1.5ex}
\begin{equation}\label{e-u-a}
  u = \sqrt{4ga} \,.
\end{equation}
With this substitution, (\ref{e-kine}) becomes (\ref{e-energy}),
so~that $v(y)$ becomes the \emph{normal} speed of the cycloidal
cylinder~\textit{HPF}, i.e.~the speed of its intersection with its
orthogonal cycloid~\textit{OPB}. Now let the particle be released
from~$O$ when cylinder~\textit{HPF} crosses~$O$. If the frictionless
path is cycloid~\textit{OPB}, the particle will \emph{just} keep up
with cylinder~\textit{HPF}, and will reach~$Z$ when
cylinder~\textit{HPF} crosses~$Z$. But if the path from $O$ to~$Z$
departs from cycloid~\textit{OPB}, then for part of the path~\! the
same $v(y)$ over the same range of $y$ will \emph{not} be normal to
cylinder~\textit{HPF}, so that the particle will need to travel
\emph{further at the same speed} between successive positions of
cylinder~\textit{HPF} and will therefore fall behind it, taking more
time to reach~$Z$. Thus \emph{the path of least time
(\textbf{brachistochrone)} from $O$ to~$Z$ is an inverted cycloid
meeting its generating line at~$O$ and passing through~$Z$}.

\medskip\small

\noindent\textbf{Optical interpretation:} ~ Johann Bernoulli's
proof (1696) assumes that the path given by the law of refraction
satisfies Fermat's principle of least time.\footnote{\sf\,E.\,Hairer
\& G.~\!Wanner, \textit{Analysis by its History},
Springer, 2008, pp.\,137--8.} The present proof needs no optical
premises. \emph{But:} The traveling cycloidal cylinder~\textit{HPF} is
an admissible \emph{wavefront} in an isotropic but non-homogeneous
medium whose propagation speed~$v$ is given by~(\ref{e-energy}). Each
associated \emph{ray} path is a path of least time between successive
positions of the wavefront, hence a path of least time between any two
points on the path. In~an \emph{isotropic} medium, a path of least
time between successive positions of the wavefront is an orthogonal
trajectory\textemdash e.g.\,cycloid~\textit{OPB}. The law
of \emph{refraction} can be derived from the premise that the incident
and refracted wavefronts have a common intersection with the
interface. In~a \emph{stratified} medium, which behaves like many
parallel plane interfaces, this premise is satisfied by a wavefront
that holds its shape while traveling parallel to the strata\textemdash
like cylinder~\textit{HPF}.\, Eq.\,(\ref{e-v-u}) says that
\mbox{$v$/$\cos\phi$}\, is constant, i.e.~that \mbox{$v$/$\sin\alpha$}\,
is constant, where~{\boldmath$\color{myc}\alpha$} (in~the diagram) is the
angle between the wave-normal and the normal to the strata; this is
``Snell's law'' for a stratified medium.

\medskip

\noindent\textbf{Revision history:} ~ This article is condensed from
the same author's ``Quick proofs of properties of the cycloid''
(Zenodo: \href{https://doi.org/10.5281/zenodo.15601996}{15601996},
2025), with an added ``optical interpretation''\!. This revision
(version~2) removes a broken link from the abstract page and adds the
abstract to other pages.

\end{document}